\documentclass[%
 aip,
 amsmath,amssymb,
reprint, onecolumn 
]{article}

\usepackage{graphicx}
\usepackage{amsmath}
\usepackage{amssymb}
\usepackage{dcolumn}
\usepackage{bm}
\usepackage{color}
\usepackage[utf8]{inputenc}
\usepackage[T1]{fontenc}
\usepackage{mathptmx}
\usepackage{etoolbox}
\usepackage{algpseudocode}
\usepackage{algorithm,algorithmicx}
\usepackage{amsmath}
\usepackage{amsthm}  
\usepackage{amsfonts} 
\newtheorem{theorem}{Theorem}[section]  
\newtheorem{lemma}[theorem]{Lemma}  
\usepackage{graphicx}
\usepackage{subcaption}
\usepackage{amsthm}
\usepackage{booktabs}
\usepackage{array}
\usepackage{geometry}
\usepackage{multirow}
\usepackage{authblk}

\providecommand{\keywords}[1]{\textbf{Keywords:} #1}
\makeatletter
\def\@email#1#2{%
 \endgroup
 \patchcmd{\titleblock@produce}
  {\frontmatter@RRAPformat}
  {\frontmatter@RRAPformat{\produce@RRAP{*#1\href{mailto:#2}{#2}}}\frontmatter@RRAPformat}
  {}{}
}%
\makeatother
\begin{document}


\title{Extracting Interaction Kernels for Many-Particle Systems by a Two-Phase Approach}
\author[a]{Yangxuan Shi}
\author[b]{Wuyue Yang \thanks{yangwuyue@bimsa.cn}}%
\author[a]{Liu Hong \thanks{hongliu@sysu.edu.cn}}
\affil[a]{School of Mathematics, Sun Yat-sen University, Guangzhou 510275, China}%
\affil[b]{Beijing Institute of Mathematical Sciences and Applications, Beijing, 101408, China}%

\maketitle

\begin{abstract}
This paper presents a two-phase method for learning interaction kernels of stochastic many-particle systems. After transforming stochastic trajectories of every particle into the particle density function by the kernel density estimation method, the first phase of our approach combines importance sampling with an adaptive threshold strategy to identify key terms in the kernel function, while the second phase uses the whole dataset to refine the coefficients. During the implementation of our method, the mean-field equation plays a key role in reformulating the task of extracting the interaction kernels into a learnable regression problem. We demonstrate the outstanding performance of our approach through extensive numerical examples, including interacting particle systems with a cubic potential, power-law repulsion-attraction potential, piecewise linear potential, as well as a two-dimensional radially symmetric potential. 
\end{abstract}

\section{Introduction}
A stochastic interacting particle system (IPS) is a large number of particles or agents interacting through nonlinear dynamics while being subject to stochastic noise. From microscopic molecular motions to macroscopic phenomena in artificial intelligence, interacting particle systems are ubiquitous both in nature and in social science. Traditional applications range from collective cell migration to human population dynamics, while emerging applications in large language models (LLMs) and multi-agent AI systems open up new directions for this framework\cite{xi2023rise}. In these AI systems, individual agents, whether different instances of language models or specialized AI agents, interact and collaborate through complex mechanisms to accomplish tasks, exhibiting collective behaviors quite similar to that of classical collective systems.

Mining and modeling such systems requires identifying possible interaction mechanisms between individuals. However, these interaction rules are usually unknown and need to be inferred from observed data. This belongs to the class of inverse problems, which however is particularly difficult in the current case, since the IPS involves a large number of interacting agents under the impacts of intrinsic and extrinsic stochastic effects.

In recent years, data-driven approaches have become a powerful tool for addressing these issues. For example, SINDy and its variants have been successfully applied to discover governing equations in different fields\cite{brunton2016discovering}. They identify sparse representations of nonlinear dynamical systems from data, often revealing interpretable physical laws. Data-driven methods have also been applied to parameter inference and model discovery for interacting particle systems\cite{messenger2022learning,supekar2023learning,lang2022learning}. Peng et al. discussed the potential of combining multiscale modeling with machine learning. By reviewing some typical case studies, they demonstrated the striking power of machine learning to extract macro- behaviors from micro-data\cite{peng2021multiscale}. The research by Vlachas et al. focused on learning effective models of complex systems. They successfully extracted simplified macroscopic models from microscopic simulation data by using machine learning techniques, significantly enhancing the understanding and simulation efficiency of complex systems\cite{vlachas2022multiscale}. A continuum description is essential for understanding the collective phenomena in active matters. By using a physically informed data-driven approach, Golden et al. constructed a complete mathematical model of an active nematic based on experimental data, revealing dynamics controlled by active and friction stresses with no role for elastic effects\cite{golden2023physically}.

In the above mentioned studies, the mean-field theory plays a role in analyzing the complex dynamics of IPS, providing a simplified framework to describe the macroscopic behaviors arising from microscopic interactions\cite{chate2008modeling,helbing2001traffic}. Despite its success, deriving accurate mean-field equations for interacting particle systems remains a challenging task, especially under dynamically changing environments\cite{motsch2014heterophilious}.

In this paper, we propose a two-phase method for learning the interaction kernels of IPS. Our starting point is the stochastic trajectories of every particle in the IPS, which are obtained either by experimental measurements or by numerical simulations. Then using the kernel density estimation approach, the empirical distribution function is constructed, which is supposed to be the solution to the corresponding mean-field equation. The interaction kernel function constitutes an essential part of the mean-field equation and determines the collective dynamics of the IPS. Now we apply importance sampling with an adaptive threshold strategy to identify the key terms in the interaction kernel, which later will be fine-tuned by making use of the whole dataset. This is the key steps of our method. 

To provide a detailed explanation on the algorithms of our method, as well as applications to several concrete examples, the remainder of the paper is structured as follows: Section 2 establishes the theoretical foundation for the mean-field equations and their connections to the stochastic interacting particle systems. Section 3 presents our Two-Phase methodology for learning interaction kernels in detail. In Section 4, we test our approach through a series of numerical experiments. The last section concludes with a summary of the findings and potential future directions.

\section{Mean-field theory for interacting particle systems}
Here we focus on how to derive many-body mean-field equations from the individual stochastic dynamics of interacting particles. Consider the following system of $N$ interacting particles (IPS),
\begin{equation}
\label{IPS}
d\mathbf{X}_{i}^{N}=\frac{1}{N}\sum_{j=1}^{N}\mathbf{K}(\mathbf{X}_{i}^{N},\mathbf{X}_{j}^{N})dt+\frac{1}{N}\sum_{j=1}^{N}\mathbf{\sigma}(\mathbf{X}_{i}^{N},\mathbf{X}_{j}^{N})\cdot d\mathbf{W}_t^i 
\end{equation}
where $\mathbf{X}_i^{N}=\mathbf{X}_i^{N}(t)\in \mathbb{R}^{d}$  represents the position of the $i$'th particle in a d-dimensional space at time $t$. $\mathbf{K}:\mathbb{R}^{d}\times\mathbb{R}^{d}\rightarrow\mathbb{R}^{d}$ denotes the drift term, $\mathbf{\sigma}:\mathbb{R}^{d}\times\mathbb{R}^{d}\rightarrow\mathbb{R}^{d}\times\mathbb{R}^{d}$ represents the diffusion term, and $\mathbf{W}_t^i$ ($1\leq i\leq N$) are $N$ independent $d$-dimensional Wiener processes.

We aim to derive the mean-field limit of this interacting particle system as $N\rightarrow\infty$. Introduce the empirical distribution function
\begin{equation}
u^N(\mathbf{x},t)\equiv\frac{1}{N}\sum_{j=1}^{N}\delta(\mathbf{x}-\mathbf{X}_j^{N}).
\end{equation}
Then, by the strong law of large numbers, we have
\begin{equation}
u(\mathbf{x},t)=\lim_{N\rightarrow\infty}u^N(\mathbf{x},t)
\end{equation}
as the density function of the particles at position $\mathbf{x}$ and time $t$. Similarly, it can be shown that 
\begin{eqnarray}
&\frac{1}{N}\sum_{j=1}^{N}\mathbf{K}(\cdot,\mathbf{X}_{j}^{N}(t))\xrightarrow{N\to\infty}\int \mathbf{K}(\cdot,\mathbf{x})u(\mathbf{x},t)d\mathbf{x},\\
&\frac{1}{N}\sum_{j=1}^{N}\mathbf{\sigma}(\cdot,\mathbf{X}_{j}^{N}(t))\xrightarrow{N\to\infty}\int \mathbf{\sigma}(\cdot,\mathbf{x})u(\mathbf{x},t)d\mathbf{x}.
\end{eqnarray}
Substituting these formulas into the SDE for the IPS, we obtain the so-called McKean-Vlasov SDE (MVE)\cite{mckean1966class},
\begin{equation}
d\overline{\mathbf{X}}_i(t)=\left[\int_{\mathbb{R}^d}\mathbf{K}(\overline{\mathbf{X}}_i(t),\mathbf{y})u(\mathbf{y},t)d\mathbf{y}\right]dt+\left[\int_{\mathbb{R}^d}\mathbf{\sigma}(\overline{\mathbf{X}}_i(t),y)u(\mathbf{y},t)d\mathbf{y}\right]\cdot d\mathbf{W}_t^i.
\end{equation}

Based on the coupling method, it becomes possible to establish rigorous convergence results between the solutions of IPS and MVE in mathematics \cite{sznitman1991topics}.
\begin{lemma}
For $\mathbf{K}, \mathbf{\sigma} \in W^{1,\infty}\cap L^{\infty}$, there exists a constant $c > 0$, such that
\begin{equation}
\mathbb{E}\left[\max_{0 \leq s \leq t}|\mathbf{X}_{i}^{N}(s)-\overline{\mathbf{X}}_i(s)|^{2}\right] \leq \frac{c}{N}e^{ct},
\end{equation}
where $\mathbf{X}_{i}^{N}(s)$ and $\overline{\mathbf{X}}_i(s)$ represent the solutions to IPS and MVE respectively.
\end{lemma}
\begin{theorem}
As a consequence, we have
\begin{equation}
(\mathbf{X}_{i}^{N}(t))_{t\in[0,T]} \xrightarrow{N\to\infty} (\overline{\mathbf{X}}_i(t))_{t\in[0,T]}\; \text{ in probability}.
\end{equation}
\end{theorem}

Let $g(\mathbf{x})\in\mathbb{C}^2(\mathbb{R}^d)$ be the test function. Applying Itô's formula and taking expectations on both sides, the MVE yields
\begin{equation}\label{ito}
\begin{aligned}
&\mathbb{E}\left[  g(\overline{\mathbf{X}}_i(t))\right]-\mathbb{E}\left[g(\overline{\mathbf{X}}_i(0))\right]=\mathbb{E}\left[\int_0^tg^{\prime}(\overline{\mathbf{X}}_i(s))\cdot\int_{\mathbb{R}^d}\mathbf{K}(\overline{\mathbf{X}}_i(s),\mathbf{y})u(\mathbf{y},s)d\mathbf{y}ds\right]\\&+\frac{1}{2}\mathbb{E}\left[\int_0^t g^{\prime\prime}(\overline{\mathbf{X}}_i(s)):\left(\int_{\mathbb{R}^d}\mathbf{\sigma}(\overline{\mathbf{X}}_i(s),\mathbf{y})u(\mathbf{y},s)d\mathbf{y}\right)^T\cdot\left(\int_{\mathbb{R}^d}\mathbf{\sigma}(\overline{\mathbf{X}}_i(s),\mathbf{y})u(\mathbf{y},s)d\mathbf{y}\right)ds\right].    
\end{aligned}
\end{equation}
Taking an average of the $N$ equations in \eqref{ito}, we obtain
\begin{equation}
\begin{aligned}
&\int_{\mathbb{R}^d} g(\mathbf{x})u(\mathbf{x},t)d\mathbf{x}-\int_{\mathbb{R}^{d}}g(\mathbf{x})u(\mathbf{x},0)d\mathbf{x}=\int_{\mathbb{R}^{d}}\left[\int_0^tg^{\prime}(\mathbf{x})\cdot\left(\int_{\mathbb{R}^{d}} K(\mathbf{x},\mathbf{y})u(\mathbf{y},s)d\mathbf{y}\right)ds\right]u(\mathbf{x},t)d\mathbf{x}\\
&+\frac{1}{2}\int_{\mathbb{R}^{d}}\left(\int_{0}^{t}g^{\prime\prime}(\mathbf{x}):\left(\int_{\mathbb{R}^{d}}\sigma(\mathbf{x},\mathbf{y})u(\mathbf{y},s)d\mathbf{y}\right)^T\cdot\left(\int_{\mathbb{R}^{d}}\sigma(\mathbf{x},\mathbf{y})u(\mathbf{y},s)d\mathbf{y}\right)ds\right)u(\mathbf{x},t)d\mathbf{x}.
\end{aligned}
\end{equation}
Using Fubini's theorem to interchange the order of integration, we differentiate above equation by $t$, which gives
\begin{equation}
\begin{aligned}
&\int_{\mathbb{R}^d}g(\mathbf{x})\frac \partial{\partial t}u(\mathbf{x},t)d\mathbf{x}=\int_{\mathbb{R}^d}g^{\prime}(\mathbf{x})\cdot\left(\int_{\mathbb{R}^d}\mathbf{K}(\mathbf{x},\mathbf{y})u(\mathbf{y},t)d\mathbf{y}\right)u(\mathbf{x},t)d\mathbf{x}\\
&+\frac{1}{2}\int_{\mathbb{R}^d}g^{\prime\prime}(\mathbf{x}):\left(\int_{\mathbb{R}^d}\mathbf{\sigma}(\mathbf{x},\mathbf{y})u(\mathbf{y},t)d\mathbf{y}\right)^T\cdot\left(\int_{\mathbb{R}^d}\mathbf{\sigma}(\mathbf{x},\mathbf{y})u(\mathbf{y},t)d\mathbf{y}\right)u(\mathbf{x},t)d\mathbf{x}\\
&=-\int_{\mathbb{R}^d}g(\mathbf{x})\nabla_x \cdot \left(u(\mathbf{x},t)\int_{\mathbb{R}^d}\mathbf{K}(\mathbf{x},\mathbf{y})u(\mathbf{y},t)d\mathbf{y}\right)d\mathbf{x}\\
&+\frac{1}{2}\int_{\mathbb{R}^d}g(\mathbf{x})\Delta_x:\left[u(\mathbf{x},t)\left(\int_{\mathbb{R}^d}\mathbf{\sigma}(\mathbf{x},\mathbf{y})u(\mathbf{y},t)d\mathbf{y}\right)^T\cdot\left(\int_{\mathbb{R}^d}\mathbf{\sigma}(\mathbf{x},\mathbf{y})u(\mathbf{y},t)d\mathbf{y}\right)\right]d\mathbf{x}.
\end{aligned}
\end{equation}
The last equality holds by integration by parts. Due to the arbitrariness of the test function, the mean-field equation reads
\begin{equation}
\label{1stMFD}
\frac{\partial}{\partial t}u(\mathbf{x},t)+\nabla_x \cdot \left[u(\mathbf{x},t)\left(\mathbf{K}\ast u(\mathbf{x},t)\right)\right]=\frac{1}{2}\Delta_{x}:\left[u(\mathbf{x},t)\left(\mathbf{\sigma}\ast u(\mathbf{x},t)\right)^T\cdot\left(\mathbf{\sigma}\ast u(\mathbf{x},t)\right)\right]
\end{equation}
where $\mathbf{K}\ast u(\mathbf{x},t)=\int_{\mathbb{R}^{d}}\mathbf{K}(\mathbf{x},\mathbf{y})u(\mathbf{y},t)d\mathbf{y}$ and $\mathbf{\sigma}\ast u(\mathbf{x},t)=\int_{\mathbb{R}^{d}}\mathbf{\sigma}(\mathbf{x},\mathbf{y})u(\mathbf{y},t)d\mathbf{y}$ represent convolution operations. Eq.\eqref{1stMFD} is also known as the McKean-Vlasov equation. Particularly, when the diffusion term $\mathbf{\sigma}(\mathbf{X}_i^N, \mathbf{X}_j^N)=\sigma_c\mathbf{I}_{d\times d}$ is a constant matrix, we have 
\begin{equation}
\label{MVE}
\frac{\partial}{\partial t}u(\mathbf{x},t)+\nabla_x \cdot \left[u(\mathbf{x},t)\left(\mathbf{K}\ast u(\mathbf{x},t)\right)\right]=\frac{\sigma_c^2}{2}\Delta_{x}u(\mathbf{x},t).
\end{equation}
This is the main equation we will consider in the following sections.

During the above procedure, we have derived the mean-field equation \eqref{1stMFD} in a formal way. It becomes crucial to establish the conditions under which the empirical measure of the IPS converges to its mean-field limit. This is given through the following statement according to Kac\cite{}. 
\begin{theorem}[Propagation of Chaos]
The particles become asymptotically independent as $N \to \infty$. Specifically, for any fixed $k \in \mathbb{N}^+$, we have
\begin{equation}
(\mathbf{X}_1^N(t),\ldots,\mathbf{X}_k^N(t)) \xrightarrow{N\rightarrow\infty} u(\mathbf{x},t)^{\otimes k}\; \text{ in distribution},
\end{equation}
where $\mathbf{X}_i^N(t)$ is the solution to IPS, $u(\mathbf{x},t)$ is the probability density solving Eq.\eqref{1stMFD}, and $u(\mathbf{x},t)^{\otimes k}$ denotes the direct products of $k$ probability density functions.
\end{theorem}

\section{A Two-Phase Method for Learning Interaction Kernels}
With respect to the McKean-Vlasov equation in \eqref{MVE}, our primary goal is to extract the explicit form of the drift term $\mathbf{K}=\mathbf{K}(\mathbf{x},\mathbf{y})$ based on a knowledge of stochastic trajectories of every particles. Here we restrict ourselves to a specific form of $\mathbf{K}$, which is supposed to be the gradient of a radial interaction potential $\Phi$, 
\begin{equation}
\mathbf{K}(\mathbf{x},\mathbf{y})=\nabla\Phi(||\mathbf{x}-\mathbf{y}||_2)=\phi(||\mathbf{x}-\mathbf{y}||_2)\frac{\mathbf{x}-\mathbf{y}}{||\mathbf{x}-\mathbf{y}||_2}.   
\end{equation}
Here $\phi(r)=\Phi'(r)$ is called the interaction kernel, in which $r=||\mathbf{x}-\mathbf{y}||_2$ represents the distance between two particles at positions $\mathbf{x}$ and $\mathbf{y}$.

Our learning framework consists of two phases. In Phase \uppercase\expandafter{\romannumeral 1}, by combining importance sampling with an adaptive threshold strategy, we can identify the key terms in the kernel function through sparse regression. In Phase \uppercase\expandafter{\romannumeral 2}, we fix the equation structure and utilize the complete dataset to perform fine tuning of the model parameters, thereby obtaining the final kernel function. This two-phase design effectively balances model sparsity and accuracy while maintaining high computational efficiency. 

\subsection{Phase \uppercase\expandafter{\romannumeral 1}: Sparsification with Importance Sampling}
\subsubsection{Dictionary Representation of Interaction Kernels}
In order to get the explicit form of $\phi(r)$, we will expand it in terms of basis functions,
\begin{equation} 
\phi(r)\approx \tilde{\phi}(r;\mathbf{\zeta}) = \sum_{k=1}^{M} \zeta_{k} \psi_{k}(r),
\end{equation}
where $\{\psi_{1}(r),\cdots,\psi_{M}\}$ are basis functions we chose and $\mathbf{\zeta}=\{\zeta_{1}(r),\cdots,\zeta_{M}\}$ are coefficients to be learned. Here we consider two different forms of basis functions. The first is a polynomial basis,
$$
\mathcal{D}_{\text{poly}} = \{r^{p_1},r^{p_2},\cdots| p_1,p_2,\cdots \in \mathbb{Z}\}.
$$
Physically, the positive powers $(p_i > 0)$ represent long-range interactions, negative powers $(p_i < 0)$ represents short-range interactions, while the case $p_i = 0$ represents constant interactions independent of distance. The second is a physics-informed dictionary basis
$$
\mathcal{D}_{\text{phys}} = \{\sin(r), \cos(r), e^r, e^{-r}, e^{-r}/r, \text{sgn}(r-r_c)r^{-1},\cdots\}
$$
which incorporates various common interactions in physics. 

We define an objective function that measures the discrepancy between the observed system evolution and predictions based on kernel function estimation
\begin{equation}
    \min_{\mathbf{\zeta}} \left| \left|\frac{\partial u}{\partial t} + \nabla_x \cdot \left(u(\tilde{\mathbf{K}}(\cdot;{\zeta})\ast u)\right) - \frac{\sigma^2_c}{2}\Delta_x u\right|\right|_2^2 + \lambda||\mathbf{\zeta}||_1
\end{equation}
where $\tilde{\mathbf{K}}(\mathbf{x},\mathbf{y};{\zeta})=\tilde{\phi}(r;\mathbf{\zeta})\frac{\mathbf{x}-\mathbf{y}}{r}$ with $r=||\mathbf{x}-\mathbf{y}||_2$, $\mathbf{\zeta}$ are coefficients for the interaction kernel $\phi(r)$, $\lambda$ is the regularization parameter, and 
$||\cdot||_1$ denotes the $L_1$ norm for sparsity. The spatial derivatives are approximated by using centered finite differences
$$
\Delta_x u\approx \frac{u(\mathbf{x}+\Delta\mathbf{x},t) - 2u(\mathbf{x},t) + u(\mathbf{x}-\Delta\mathbf{x},t)}{(\Delta\mathbf{x})^2}.
$$
The temporal derivative is approximated by using forward differences
$$
\frac{\partial u}{\partial t} \approx \frac{u(\mathbf{x},t+\Delta t)-u(\mathbf{x},t)}{\Delta t}.
$$
And the interaction term is computed using discrete convolution
$$
(\tilde{\mathbf{K}}(\cdot;\mathbf{\zeta})\ast u)(\mathbf{x},t)\approx \sum_{j=1}^{\infty}\sum_{||\mathbf{y}-\mathbf{x}||_2=j\Delta x} \tilde{\phi}(j\Delta x;\mathbf{\zeta})u(\mathbf{y},t)\frac{\mathbf{x}-\mathbf{y}}{j}.
$$

\subsubsection{Importance Sampling and Sparse Regression}
In the first phase, we need to correctly identify the interaction kernel and improve the sparsity of the coefficients. To achieve this goal, we use an iterative optimization process combined with important sampling to effectively explore the time domain. Specifically, instead of optimizing with respect to all available time steps, we only sample a subset of time steps according to their relative importance, where the sampling probability $p_n$ for each time step $n$ is determined by the following formula
\begin{equation}
p_n = \frac{E_n}{\sum_{k=1}^{N_t} E_k},
\end{equation}
where $E_n$ indicates the error of the $n$'th time step. This method ensures that those time steps with large errors will get more chances to be sampled during the optimization procedure. 

To improve the sparsity of coefficients, we implement a mechanism of adaptive thresholds. During the $k$'th iteration, the threshold $\tau_k$ is determined by
\begin{equation}
\tau_k = \alpha\max \{|\zeta_{1}^{(k)}|,\cdots, |\zeta_{M}^{(k)}|\},
\end{equation}
where $\alpha$ is the sparsity intensity and $\zeta_{i}^{(k)}$ is the $i$'th coefficient during the $k$'th iteration. Record a binary mask matrix $\mathbf{m} \in \{0,1\}^{M}$ , which will be updated according to the following rule on threshold
\begin{equation}
m_{i}^{(k+1)} =\left\{
\begin{array}{l}
1,\; \text{if}\; |\zeta_{i}^{(k)}| \geq \tau_k;\\
0,\; \text{if}\; |\zeta_{i}^{(k)}| < \tau_k.
\end{array}
\right.
\end{equation}

\subsection{Phase \uppercase\expandafter{\romannumeral 2} : Refinement with Frozen Structure}
The second phase focuses on refining the coefficients of the identified kernel function in Phase I while still maintaining its sparsity (keeping the mask matrix unchanged). Unlike Phase I, this phase is optimized using all available time steps, with the objective function set as
\begin{equation}
\min_{\mathbf{\zeta}} \sum_{n=1}^{N_t}\left| \left|\frac{\partial u}{\partial t}(\cdot,t_n) + \nabla_x \cdot \left(u(\cdot,t_n)(\tilde{\mathbf{K}}(\cdot;\mathbf{\zeta}\odot\textbf{m})\ast u)\right) - \frac{\sigma^2_c}{2}\Delta_x u(\cdot,t_n)\right|\right|_2^2 + \lambda||\mathbf{\zeta}\odot\textbf{m}||_1
\end{equation}
where $t_n$ represents the $n$'th time step, $\mathbf{m}$ is the binary mask matrix determined in Phase \uppercase\expandafter{\romannumeral 1}, and $\mathbf{\zeta}\odot\textbf{m}=(\zeta_{1}m_{1},\cdots,\zeta_{M}m_{M})$ represents sparse coefficients by masking. The detailed implementation steps including Phase I and Phase II are provided in Algorithm \ref{algorithm}.



\begin{algorithm}
\caption{A Two-Phase Method for Learning Interaction Kernels with Importance Sampling and Sparsity Frozen.}
\label{algorithm}
\begin{algorithmic}[1]
	\Require Total number of particles $N$, time interval $dt$, space grid $\mathbf{x}$ 
	\Ensure Optimized coefficients $\boldsymbol{\zeta}$ for the kernel function \\
	\textit{\#\textbf{Data Generation:} Generate particle trajectories through SDEs}
	\State Initialize the $N$ particle system $\{\mathbf{X}_i(0)\}_{i=1}^N$
	\For{each time step $t$}
	\State Update particle positions $\{\mathbf{X}_i(t)\}_{i=1}^N$ according to Eq. \eqref{IPS}
	\State Store particle positions for each time step $t$
	\EndFor
	\State Use KDE to estimate particle density function $u(\mathbf{x}, t)$ at each time step
	\State Generate candidate basis function dictionary $\{\Psi_1(\mathbf{x}),\cdots,\Psi_M(\mathbf{x})\}$\\
	\textit{\#\textbf{Phase I:} Importance sampling with sparsity regression} 
	\State Initialize coefficients $\boldsymbol{\zeta}$ randomly
	\State Initialize the mask matrix $\mathbf{m}$ as a boolean array with all true values
	\Repeat
	\State Compute time derivative $\partial_t u$ and space derivative $\partial_x u$
	\State Compute error $E_n$ at each time step
	\State Calculate sampling probability: $p_n = E_n/\sum_{k=1}^{N_t} E_k$
	\State Sample time steps according to $p_n$
	\State Compute adaptive thresholds $\tau_k = \alpha\max\{|\zeta_{1}^{(k)}|,\cdots,|\zeta_{M}^{(k)}|\}$
	\State Update coefficients $\boldsymbol{\zeta}$ by using sampled time steps
	\State Update the mask matrix: $m_{i}^{(k+1)} = \mathbf{1}(|\zeta_{i}^{(k)}| \geq \tau_k)\; \text{or}\; \mathbf{0}(|\zeta_{i}^{(k)}|<\tau_k)$
	\Until{convergence or all coefficients in $\boldsymbol{\zeta}$ are masked}\\
	\textit{\# \textbf{Phase II:} Refinement with fixed mask matrix}
	\State Using fixed mask matrix obtained from Phase \uppercase\expandafter{\romannumeral 1}
	\State Minimize loss function $\sum_{n=1}^{N_t}\left| \left|\frac{\partial u}{\partial t}(\cdot,t_n) + \nabla_x \cdot \left(u(\cdot,t_n)(\tilde{\mathbf{K}}(\cdot;\mathbf{\zeta}\odot\textbf{m})\ast u)\right) - \frac{\sigma^2_c}{2}\Delta_x u(\cdot,t_n)\right|\right|_2^2 + \lambda||\mathbf{\zeta}\odot\textbf{m}||_1$ with all time steps
	\State \Return optimized $\boldsymbol{\zeta}$.
\end{algorithmic}
\end{algorithm}

\begin{figure}
\centering
\includegraphics[width=0.9\linewidth]{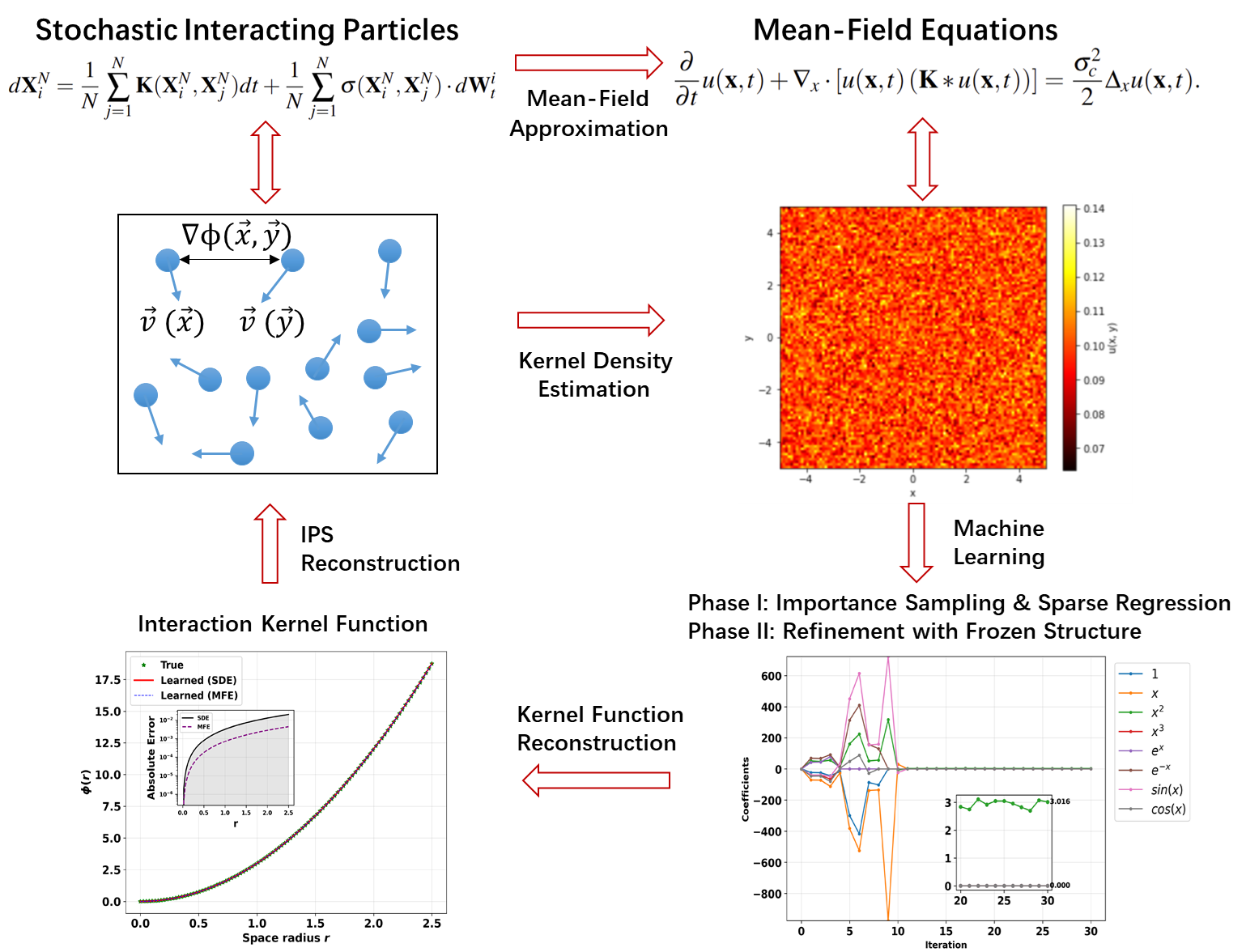}
\caption{Flowchart for interaction kernel extraction by our two-phase approach for many particle systems.}
\label{fig:case1_phi}
\end{figure}

\subsection{Results assessment}
We evaluate the results using three different criteria: comparison of the estimated values with the ground truth, the Wasserstein distance between the estimated and true solutions, and the estimated and true free energy flows.

\paragraph{Estimated and true kernels}

We compare the true and estimated kernels either by plotting their numerical values side-by-side or by calculating their relative errors on the coefficients when explicit expressions are available. Such a comparison allows us to visualize their similarities and differences.

\paragraph{Wasserstein distance}

To quantify the difference between the solutions calculated based on the true and estimated interaction kernels, we refer to the 2-Wasserstein distance between them. The 2-Wasserstein distance \( W_2(f, g) \) measures the optimal transport cost between two probability density functions \( f \) and \( g \) over the spatial domain \( \Omega \), which is defined as:
\[
W_2(f, g) := \left( \inf_{\gamma \in \Gamma(f,g)} \int_{\Omega \times \Omega} |x - y|^2 d\gamma(x, y) \right)^{1/2}
\]
where \( \Gamma(f,g) \) denotes the set of all couplings between the two densities \( f \) and \( g \). The Wasserstein distance provides a rigorous metric for assessing the accuracy of the estimated solutions compared to the true solutions, which is particularly useful when the differences are not visually apparent.

\paragraph{Free energy}

We also compare the true and estimated free energies. The free energy, whose Wasserstein gradient gives the mean-field equation, is defined by:
\[
E[u,\mathbf{\phi}](t)=\frac{\sigma_c^2}{2}\int_{\mathbb{R}^d}u\ln(u)\,d\mathbf{x} + \int_{\mathbb{R}^d}u(u\ast \Phi)\,d\mathbf{x}, \quad \text{with} \quad \Phi(r)=\int_0^r\mathbf{\phi}(s)\,ds.
\]
The true and estimated free energies are denoted as \( E[u,\phi](t) \) and \( E[\widehat u,\widehat\phi](t) \), respectively, allowing us to evaluate how well the estimated interaction kernel function captures the underlying dynamics of the IPS.

\section{Numerical Experiments}
Here we consider a stochastic system composed of $N$ particles moving in space with two-body non-local interactions among them. The dynamics of this system are described by a group of coupled SDEs in Eq. \eqref{IPS}. As we have shown, in the mean-field limit $N \to \infty$, this system converges to the mean-field equation in Eq. \eqref{MVE}. Now our primary goal is to learn the interaction kernels from the observational data of the system, the trajectories of particles to be exact. 

\subsection{Cubic potential}
In the first case, we consider particles confined in a one-dimensional tube with a cubic potential  $\phi(r) = 3r^2$ as the interaction kernel, where $r$ is the distance between any two particles. Meanwhile, the diffusion coefficient $\sigma_c$ is a constant. To perform numerical simulations, we set up a system composed of 20,000 particles, whose initial positions are sampled from the following bimodal distribution
$$\frac{1}{2}(\mathcal{N}(1,s^2) + \mathcal{N}(-1,s^2)),$$
where $\mathcal{N}$ denotes the normal distribution and $s^2$ is the variance of each normal distribution. Here we set $s^2=0.25$ 
and $\sigma_c^2=2$. Then, by using the Euler–Maruyama scheme, the IPS described by Eq. \eqref{IPS} is simulated over the time interval $[0, 1]$  with a time step $dt = 10^{-4}$. To estimate the particle density $u(x,t)$, we adopt the Kernel Density Estimation (KDE) on a spatial grid within the range $[-4, 4]$, discretized into 100 intervals.

On the other hand, we solve the mean-field equation numerically by using the fourth-order Runge-Kutta scheme to calculate time derivatives and the forward Euler scheme for spatial derivatives.  The mean-field equation is discretized on the spatial domain $[-4, 4]$ with a mesh size $dx = 0.05$ and the temporal domain $[0,1]$ with a time step $dt = 10^{-3}$.

\begin{figure}
\centering
\includegraphics[width=1.0\linewidth]{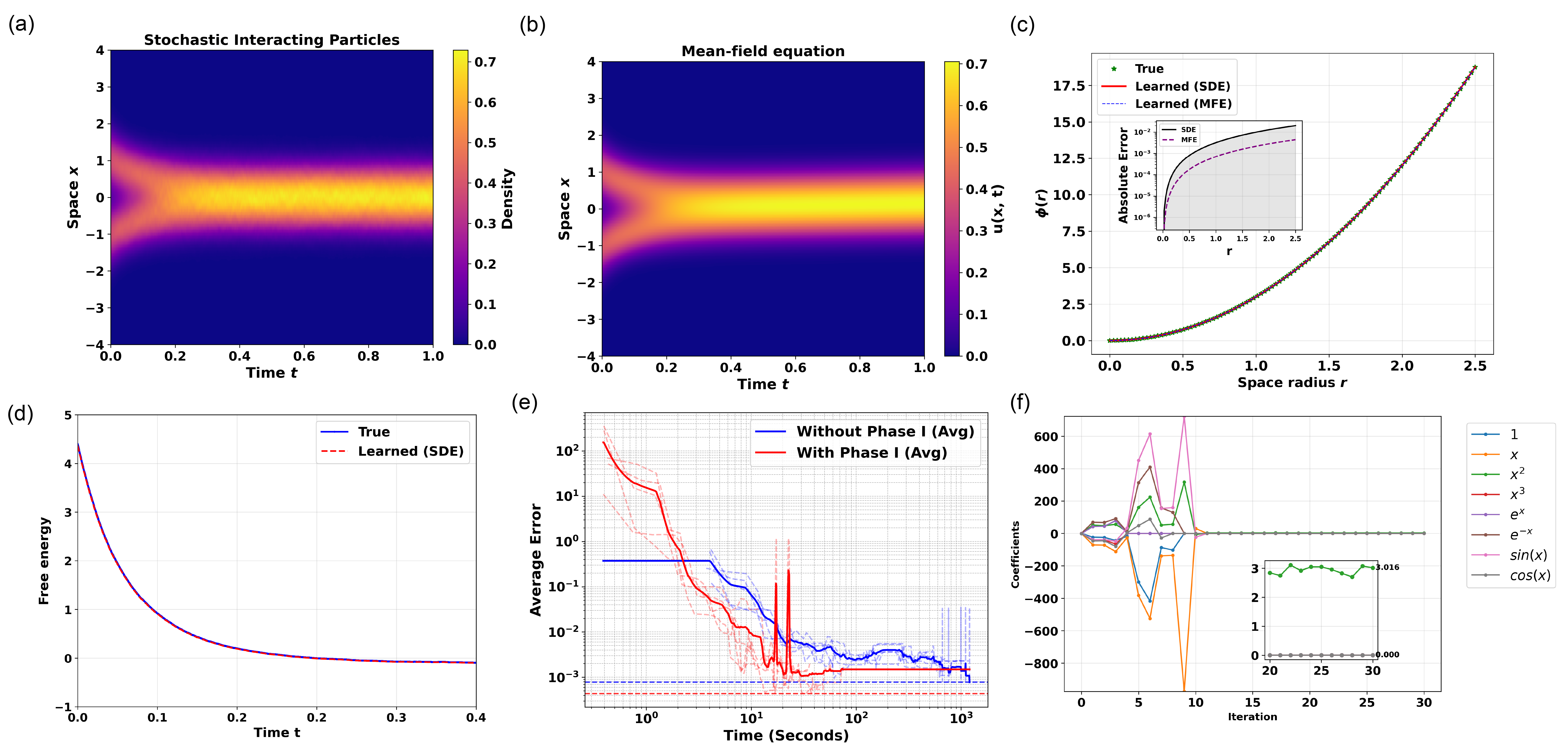}
\caption{\textbf{Identification of interaction kernels using the two-phase method in the case of cubic potential.} Evolution of particle density from $t=0$ to $t=1$, calculated by (a) coupled stochastic differential equations and (b) the mean-field equation, respectively. (c) The learned interaction kernel $\phi(r)$ from either stochastic trajectories or the particle density with an inset showing their absolute errors. (d) Comparison on the free energy for both true and learned dynamics. (e) Convergence rates with and without importance sampling in Phase I. (f) Coefficients evolution during the procedure of sparse regression.}
\label{fig:case1_1}
\end{figure}

\begin{figure}
\centering
\includegraphics[width=0.7\linewidth]{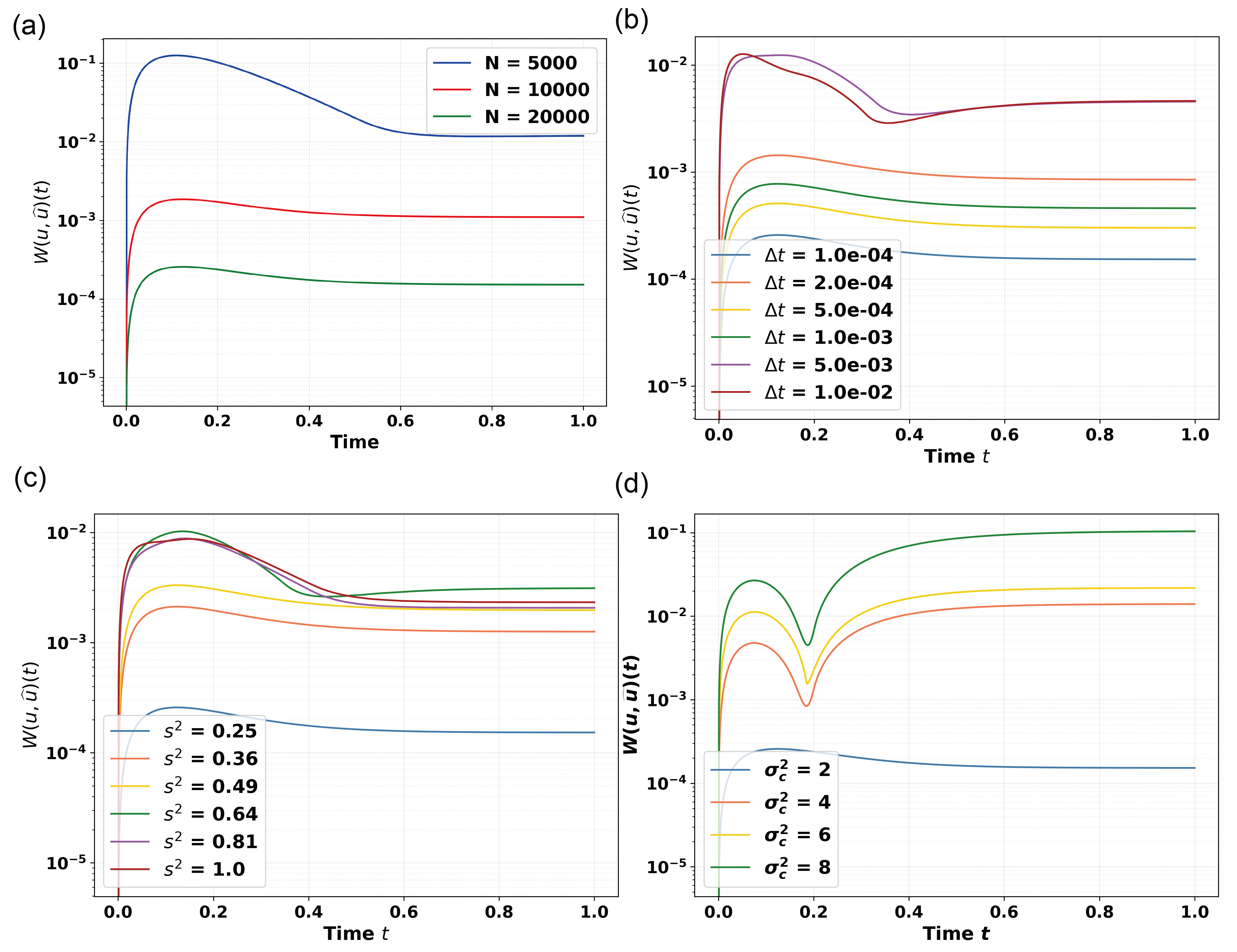}
\caption{\textbf{Comparison on the predicted particle density $\hat{u}(\mathbf{x},t)$ and the exact one $u(\mathbf{x},t)$ based on their Wasserstein distance $W_2(u, \hat{u})$.} (a) The total number of interacting particles $N$, (b) time step size $\Delta t$, (c) variance $s^2$, and (d) diffusion coefficient $\sigma_c^2$ are varied respectively.}
\label{fig:case1_2}
\end{figure}

Through comparing Fig.\ref{fig:case1_1}(a) and Fig.\ref{fig:case1_1}(b), we can see an excellent agreement on the results of stochastic particle simulations and the mean-field equation, as long as the particle number is large enough. The performance of our two-phase method is demonstrated through a direct comparison on both the form of the interaction kernel $\phi(r)$ in Fig.\ref{fig:case1_1}(c) and the free energy in Fig.\ref{fig:case1_1}(d). The inset shows that the absolute errors of the learned interaction kernel function remain consistently below $10^{-2}$. The significance of Phase I is highlighted in Fig.\ref{fig:case1_1}(e), which reveals the adoption of importance sampling could effectively reduce the computation time by an order of magnitude. The procedure of sparsity regression is drawn in the last plot. After about ten rounds of iteration, it can seen that only the coefficient for $x^2$ term remains significant and fluctuates around its exact value 3, matching the true cubic potential. 

The robustness of our method is examined through the Wasserstein distance under various parameters in Fig.\ref{fig:case1_2}. By increasing the particle number $N = 5000$, $10000$, and $20000$, the accuracy of our method is improved by above two orders of magnitude, which can be attributed to the discrepancy between the IPS and its mean-field limit. In our approach, we implement a multi-scale framework by downsampling the data in space and time to generate a coarser mesh. The coarser time step is defined as $\Delta t = k dt$, where $k \in \{1, 2, 5, 10, 50, 100\}$. As shown in Fig.\ref{fig:case1_2}(b), smaller time steps (e.g. $\Delta t = 10^{-4}$) are more accurate with a Wasserstein distance around $10^{-4}$, while larger time steps (e.g. $\Delta t = 10^{-2}$) result in higher errors around $10^{-2}$. We also explore the impacts of the initial variance $s^2$ in the bimodal distribution, with values ranging from $0.25$ to $1.0$, and the diffusion coefficient $\sigma_c^2$ in Fig.\ref{fig:case1_2}(c-d), both showing that smaller values of variance and diffusion coefficient will lead to better predictions.

\begin{table}[ht]
    \centering
    \caption{Summary on the learned interaction kernel functions.}
    \begin{tabular}{@{}ll*{9}{>{\centering\arraybackslash}p{1.2cm}}@{}}
        \toprule
        & & $x^0$ & $x^1$ & $x^2$ & $x^3$ & $e^x$ & $e^{x^0}$ & $\sin(x)$ & $\cos(x)$ & $x^{-p}$ \\
        \midrule
        \multirow{2}{*}{Case A} & True    & 0 & 0 & 3    & 0 & 0 & 0 & 0 & 0 & --\\
                                & Learned & 0 & 0 & 3.003 & 0 & 0 & 0 & 0 & 0 & --\\
        \midrule
        \multirow{2}{*}{Case B} & True    & 0 & 1 & 0    & 0 & 0 & 0 & 0 & 0 & -1 \\
                                 & Learned & 0 & 0.986 & 0    & 0 & 0 & 0 & 0 & 0 & -0.815 \\
        \midrule
        \multirow{2}{*}{Case D} & True    & 0 & 0 & 3    & 0 & 0 & 0 & 0 & 0 & -- \\
                                 & Learned & 0 & 0 & 2.916    & 0 & 0 & 0 & 0 & 0 & -- \\
        \bottomrule
    \end{tabular}
    \label{tab:candidate_basis}
\end{table}

\subsection{Power-law repulsion-attraction potential}
A classic example of a long-range interaction in physics is the Coulomb's law\cite{huray2009maxwell}, which states that the electrostatic force between two charged particles is proportional to {$r^{-2}$} . Inspired by this physical law, we consider the following interaction potential
$$\phi(r) = r - r^{-1.5},$$
where the first term $r$ represents linear long-range attraction and the second term $r^{-1.5}$ becomes singular as $r \to 0$. To address this problem, we introduce a modified potential with a cutoff radius $r_c$
\begin{equation*}
\phi(r) = \begin{cases}
r - r^{-1.5}, & r > r_c \\
r - r_c^{-1.5}, & r \leq r_c.
\end{cases}
\end{equation*}
In this case, we set the cutoff radius $r_c = 0.05$.

\begin{figure}
\centering
\includegraphics[width=1.0\linewidth]{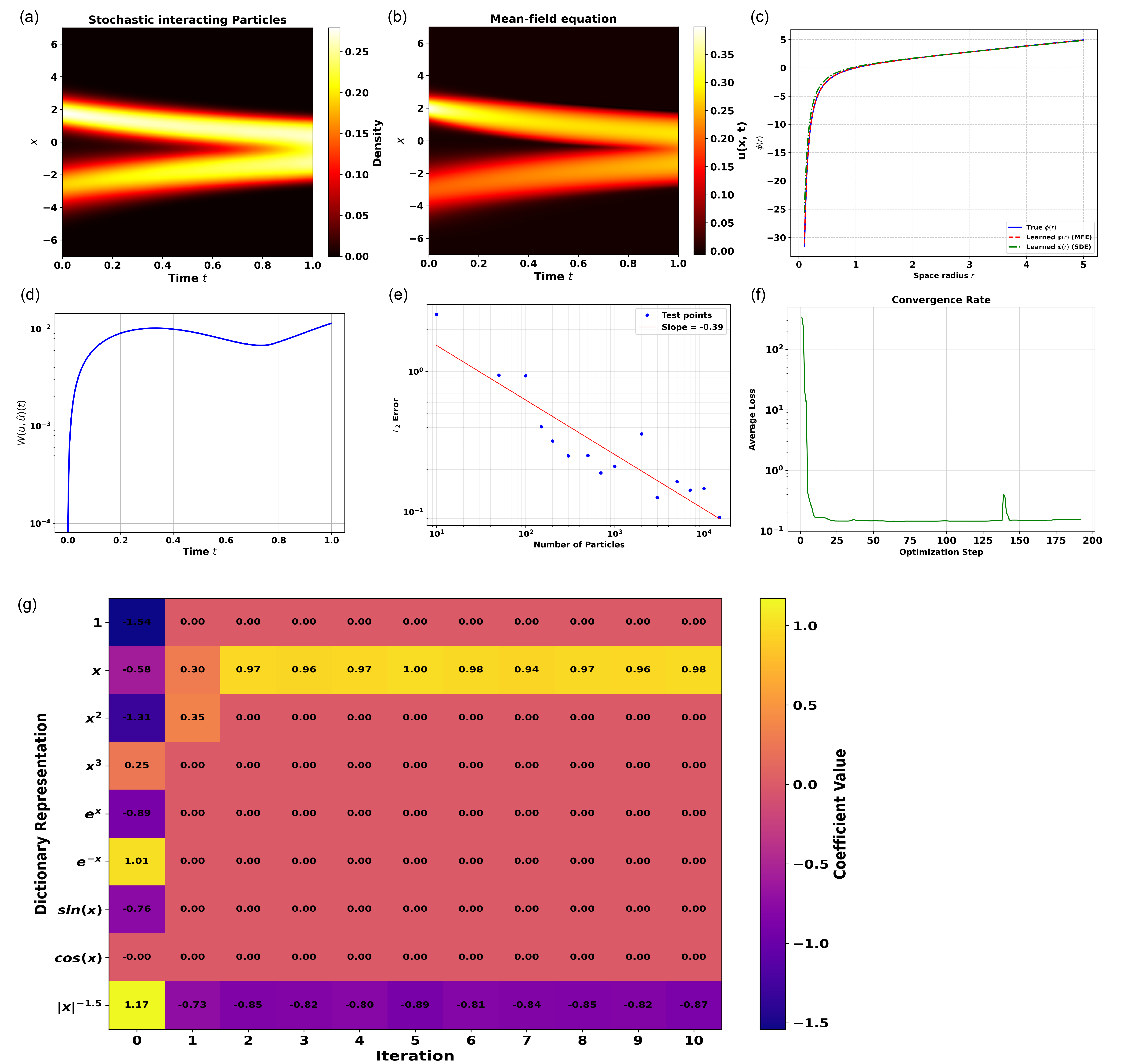}
\caption{\textbf{Results for the power-law repulsion-attraction potential.} Evolution of particle density calculated by the (a) SDE and (b) the mean-field equation. (c) Comparison between true and learned interaction kernels $\phi(r)$. (d) Wasserstein distance $W_2(u, \hat{u})$ between true and learned solutions over time. (e) The scaling relationship between the $L_2$ error and particle numbers. (f) The average loss as a function of the optimization steps, with an inset displaying the convergence rate. (g) The procedure of sparse regression. }
\label{fig:case2}
\end{figure}

By implementing the above modified potential to simulations of both SDE and the mean-field equation, we can see that it correctly reproduces the attractive forces at large distances and repulsive forces at short distances, as illustrated in Fig. \ref{fig:case2}(a-c). For the SDE simulation, we use $N = 20,000$ particles and a time step $\Delta t = 10^{-4}$ up to the final time $T = 1.0$. The diffusion constant is $\sigma_c^2=0.02$; while for the numerical solvation of the mean-field equation, we choose a spatial domain $[-7, 7]$ with the mesh size $\Delta x = 0.05$ and a time region $[0,1]$ with the time step $dt = 10^{-3}$. The initial condition is given by
$$u(x,0) = \frac{1}{2}(\mathcal{N}(2,0.25) + \mathcal{N}(-3,1)).$$

The Wasserstein distance between the estimated and true density distributions in Fig.\ref{fig:case2}(d) demonstrates that our Two-Phase method could maintain low errors below $10^{-2}$. Meanwhile, the average loss function, defined as 
$$L_{data}=\frac{1}{N_tN_x}\left|\left|u(\mathbf{x},t)-\hat{u}(\mathbf{x},t)\right|\right|_2^2,$$
converges after 20 rounds of iteration in the current case. Here, $N_t\times N_x$ denote the total number of time-space points used for evaluation, $u$ and $\hat{u}$ represent the true and estimated particle density functions, respectively. Furthermore, the increase of particle numbers could help to reduce the error between the estimated and true particle density functions, which highlights the key role of an accurate density function for the following procedure of interaction kernel extraction. Data in Fig.\ref{fig:case2}(e) gives $L_{data}\sim N^{-0.39}$. 

\subsection{Opinion dynamics with discontinuities}
\begin{figure}
\centering
\includegraphics[width=1\linewidth]{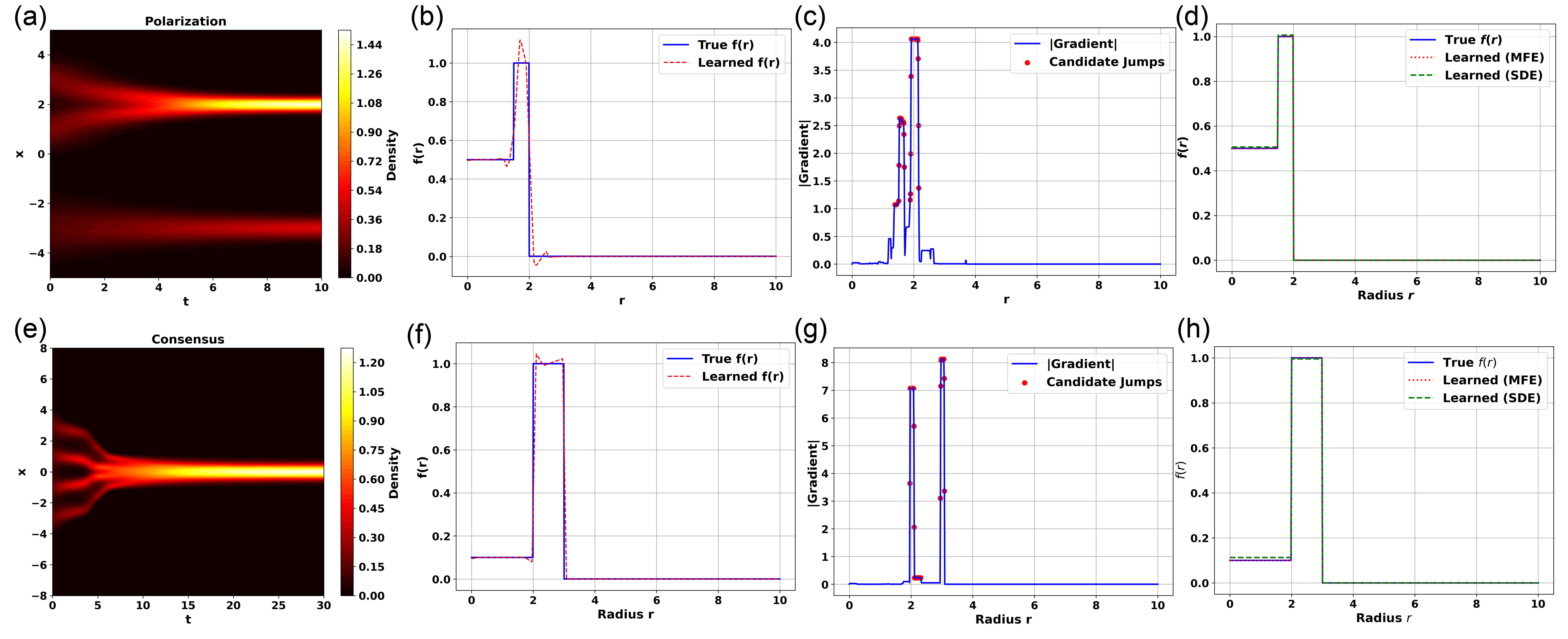}
\caption{\textbf{Opinion dynamics with discontinuities.} Evolution of opinion density distribution showing (a) polarization and (e) consensus. (b,f) Comparison between true and learned interaction functions by using DNN. (c,g) Gradient of learned interaction functions $f(r)$. (d,h) Comparison between true and learned interaction functions $f(r)$ by applying our approach.}
\label{fig:OD_1}
\end{figure}

The IPS in Eq. \eqref{IPS} is also suitable for describing the spreading of opinions among people. Let $X_i \in \mathbb{R}$ be the opinion of the $i$'th agent. According to Motsch et al.\cite{motsch2014heterophilious}, we consider two types of interaction kernels $\phi(r)=f(r)r$, i.e.
\paragraph{Polarization:}
\begin{equation*}
f_1(r) = \begin{cases}
0.5, & 0 \leq r \leq 1.5 \\
1.0, & 1.5 < r \leq 2 \\
0, & r > 2,
\end{cases}
\end{equation*}
\paragraph{Consensus:}
\begin{equation*} 
f_2(r) = \begin{cases}
0.1, & 0 \leq r \leq 2 \\
1.0, & 2 < r \leq 3 \\
0, & r > 3.
\end{cases}
\end{equation*}
As shown in Fig.\ref{fig:OD_1}(a-b), for the polarization case, the dynamics reveals a two-cluster pattern, so that opinions of agents will separate into two stable clusters; while for the consensus case, initially diverse opinions converge to a single cluster, meaning agents will reach a consensus at last.

To robustly determine the exact locations of discontinuities, we take additional steps in complement to our two-phase method, which are summarized as follows: (1) Constructing the probability density function by KDE; (2) Employing Deep Neural Networks (DNN) to approximate the interaction function $f(r)$, which are optimized through minimizing the residue loss function for the mean-field equation, i.e.
\begin{equation*}
\mathcal{L}_{MFE} = \left|\left|\frac{\partial u(x,t)}{\partial t} - \frac{\sigma_c^2}{2} \frac{\partial^2 u(x,t)}{\partial x^2} - \frac{\partial}{\partial x}\big\{u(x,t) \cdot \big[\hat{\phi}(|y-x|)\cdot(y-x)\ast u(y,t)\big]\big\}\right|\right|_2^2. 
\end{equation*}
(3) Calculating the gradient of $f(r)$ by automatic differentiation of DNN, ranking these values and clustering the top 50 points to determine the most probable locations for discontinuities; (4) making symbolic regression of $f(r)$ piece by piece based on a pre-knowledge of discontinuities. During the above procedure, the determination of discontinuities plays a role and saves us from the hard task of approximating discontinuous functions as a whole. 

Actually, from Fig.\ref{fig:OD_1}(b,f), we can observe that DNN fails to capture the exact locations of discontinuities of these interaction kernels. In contrast, our two-phase approach combined with discontinuity determination by gradients not only converges significantly faster than the DNN approach, but also successfully reconstructs the discontinuous interaction kernels, specifically for cases of polarization ($I_1: x \leq 1.5$, $I_2: 1.5 < x \leq 2$, $I_3: x > 2$) and consensus ($I_1: x \leq 2$, $I_2: 2 < x \leq 3$, $I_3: x > 3$). As shown in TABLE \ref{Tab:OD}, our learned coefficients closely match their respect true values. Here $1_{I_i(x)}$ represents the indicator function for region $i$ and the terms $x \cdot 1_{I_i(x)}$ represent linear functions within each region. 0 indicates terms that were correctly eliminated through Phase I.
\begin{table}[h!]
\caption{Piecewise linear functions learned for the opinion dynamics.}
\centering
\begin{tabular}{cccccccc}
\toprule
\textbf{Opinion Dynamics} &  & $1 \cdot 1_{I_1(x)}$ & $x \cdot 1_{I_1(x)}$ & $1 \cdot 1_{I_2(x)}$ & $x \cdot 1_{I_2(x)}$ & $1 \cdot 1_{I_3(x)}$ & $x \cdot 1_{I_3(x)}$ \\
\midrule
\multirow{2}{*}{Polarization} & True & 0.5 & 0 & 1.0 & 0 & 0 & 0 \\
& Learned & 0.506 & 0 & 1.006 & 0 & 0 & 0 \\
\midrule
\multirow{2}{*}{Consensus} & True & 0.1 & 0 & 1 & 0 & 0 & 0 \\
& Learned & 0.113 & 0 & 0.996 & 0 & 0 & 0 \\
\bottomrule
\end{tabular}
\label{Tab:OD}
\end{table}

\subsection{2D radial symmetric potential}
\begin{figure}
\centering
\includegraphics[width=1.0\linewidth]{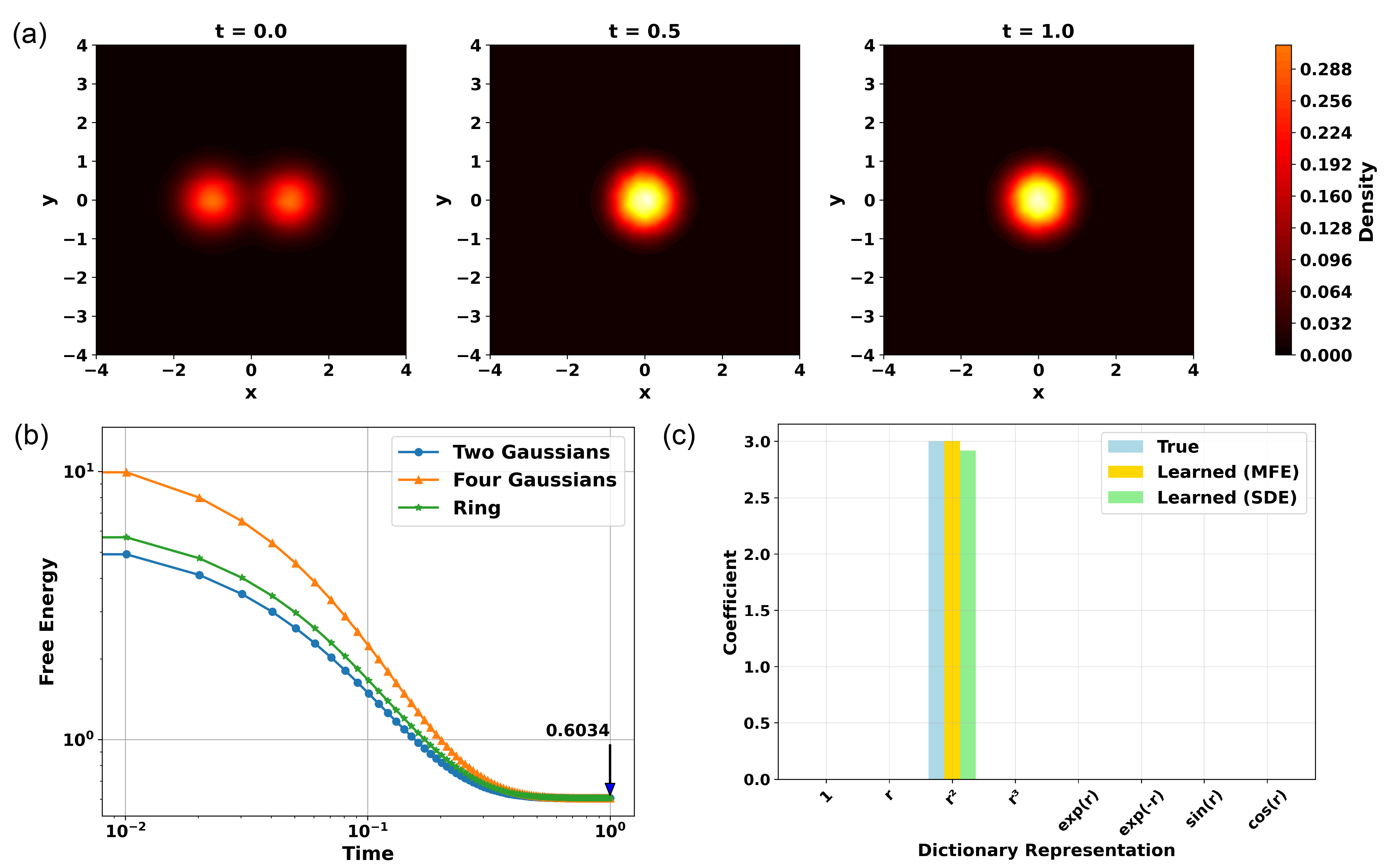}
\caption{\textbf{2D radial symmetric potential.} (a) Evolution of the density distribution estimated based on stochastic simulations of 20,000 particles. (b) Free energy under different initial conditions. (c) Comparison between true and learned coefficients in the dictionary representation of the kernel function.}
\label{fig:case3}
\end{figure}

In the last example, we extend our studies to the two-dimensional IPS. The interaction kernel $\phi(r)$ is supposed to depend only on the distance between two particles, making it radially symmetric. In particular, we set $\phi(r) = 3r^2$ and the diffusion constant $\sigma_c^2=2$. Here three initial conditions, including two Gaussians, four Gaussians, and a ring-like distribution, are explored, whose concrete forms are given by
$$
\begin{aligned}
&\text{Two Gaussians:}\quad u_1^0(x,y) = \frac{1}{2}\left(\frac{1}{2\pi s^2}\exp{\left(-\frac{(x-1)^2+y^2}{2 s^2}\right)+\frac{1}{2\pi s^2}\exp{\left(-\frac{(x+1)^2+y^2}{2 s^2}\right)}}\right),\\
&\text{Four Gaussians:}\quad u_2^0(x,y)= \frac{1}{4}\sum_{m,n=\pm1}\frac{1}{2\pi s^2}\exp\left(-\frac{(x-m)^2 + (y-n)^2}{2 s^2}\right),\\
&\text{Ring-like:}\quad u_3^0(x,y)=\frac{1}{2\pi s^2}\exp{\left(-\frac{(\sqrt{x^2+y^2}-1)^2}{2s^2}\right)},\\
\end{aligned}
$$
where $s^2 = 0.25$. The numerical results based on stochastic particle simulations are given in Fig. \ref{fig:case3}(a), which shows that the particle density function starting with a bimodal distribution will evolve towards a single peak due to the existence of long-range attractive forces. Alternative initial conditions, including four Gaussian functions and ring-like distributions, would lead to diverse two-dimensional patterns, but they all eventually converge to a single Gaussian peak at equilibrium, with free energies approaching the same value of 0.6034 (See Fig.\ref{fig:case3}(b) and Fig.S2 in SI). These diverse evolutions could be well captured by our two-phase method. And with respect to the candidate dictionary of basis functions, our approach correctly identifies the true term and its concrete coefficient. All remaining redundant terms are eliminated during the procedure of sparse regression. The result based on data from the mean-field equation is slightly better than that based on data of SDEs. See Fig. \ref{fig:case3}(c) for details.

\section{Conclusion and Discussions}
The interacting particle systems are widely known for their intrinsic complexity and fruitful phenomena. How to extract their underlying dynamics, especially the roles governing interactions between particles, is a challenging issue in the field of data-driven science. In this paper, we have developed a two-phase method to reveal the interaction kernels for many-particle systems. In Phase I, we combine importance sampling with adaptive sparsification to identify the key terms of the kernel functions, and Phase II performs refined parameter estimation with respect to the whole dataset. During the implementation of our approach, the kernel density estimation and the mean-field equation play key roles in constructing the particle density function from stochastic trajectories of interacting particles, and transforming the task of extracting the interaction kernels into a learnable linear regression problem, respectively. Through diverse numerical experiments, it has been shown that our method could successfully reconstruct various interaction kernels, including cubic potential, power-low repulsion-attraction potential, potential function with discontinuities, etc., in an accurate and robust way. 

While our proposed two-phase approach shows good performance in learning interaction kernels for low-dimensional systems, challenges arise when applied to high-dimensional systems. The major obstacle is that the kernel-based methods become less efficient with the increase of problem dimension. To address this issue, we may turn to either sample-based methods\cite{chen2021solving} or weak-form-based approaches\cite{lu2024weak}. Related works are going on. As our method based on the knowledge of empirical density function and mean-field equations, a large number of interacting particles are required to make sure the validity of the mean-field approximation, which undoubtedly puts a heavy burden on the study. How to extract the interaction kernel with respect to a small amount data of particle trajectories puts a challenge on our future studies. An alternative exciting application of our method would be the datasets of intelligent agents, similar to the foundation models of ASAL for discovering artificial life simulations\cite{kumar2024automating}. This could help us to understand the collective behaviors from the particle-based insights to intelligent-agent-based insights.

\section*
{Declaration of Competing Interest}
Authors declare that they have no conflict of interest.
\section*{Acknowledgements}
This work was supported by the National Key R\&D Program of China (2023YFC2308702), the National Natural Science Foundation of China (12301617), and Guangdong Basic and Applied Basic Research Foundation (2023A1515010157).

\section*{Reference}
\bibliographystyle{unsrt}
\bibliography{ref}

\begin{thebibliography}{10}

\bibitem{xi2023rise}
Zhiheng Xi, Wenxiang Chen, Xin Guo, Wei He, Yiwen Ding, Boyang Hong, Ming
  Zhang, Junzhe Wang, Senjie Jin, Enyu Zhou, et~al.
\newblock The rise and potential of large language model based agents: A
  survey.
\newblock {\em arXiv preprint arXiv:2309.07864}, 2023.

\bibitem{brunton2016discovering}
Steven~L Brunton, Joshua~L Proctor, and J~Nathan Kutz.
\newblock Discovering governing equations from data by sparse identification of
  nonlinear dynamical systems.
\newblock {\em Proceedings of the national academy of sciences},
  113(15):3932--3937, 2016.

\bibitem{messenger2022learning}
Daniel~A Messenger, Graycen~E Wheeler, Xuedong Liu, and David~M Bortz.
\newblock Learning anisotropic interaction rules from individual trajectories
  in a heterogeneous cellular population.
\newblock {\em Journal of the Royal Society Interface}, 19(195):20220412, 2022.

\bibitem{supekar2023learning}
Rohit Supekar, Boya Song, Alasdair Hastewell, Gary~PT Choi, Alexander Mietke,
  and J{\"o}rn Dunkel.
\newblock Learning hydrodynamic equations for active matter from particle
  simulations and experiments.
\newblock {\em Proceedings of the National Academy of Sciences},
  120(7):e2206994120, 2023.

\bibitem{lang2022learning}
Quanjun Lang and Fei Lu.
\newblock Learning interaction kernels in mean-field equations of first-order
  systems of interacting particles.
\newblock {\em SIAM Journal on Scientific Computing}, 44(1):A260--A285, 2022.

\bibitem{peng2021multiscale}
Grace~CY Peng, Mark Alber, Adrian Buganza~Tepole, William~R Cannon, Suvranu De,
  Savador Dura-Bernal, Krishna Garikipati, George Karniadakis, William~W
  Lytton, Paris Perdikaris, et~al.
\newblock Multiscale modeling meets machine learning: What can we learn?
\newblock {\em Archives of Computational Methods in Engineering},
  28:1017--1037, 2021.

\bibitem{vlachas2022multiscale}
Pantelis~R Vlachas, Georgios Arampatzis, Caroline Uhler, and Petros
  Koumoutsakos.
\newblock Multiscale simulations of complex systems by learning their effective
  dynamics.
\newblock {\em Nature Machine Intelligence}, 4(4):359--366, 2022.

\bibitem{golden2023physically}
Matthew Golden, Roman~O Grigoriev, Jyothishraj Nambisan, and Alberto
  Fernandez-Nieves.
\newblock Physically informed data-driven modeling of active nematics.
\newblock {\em Science Advances}, 9(27):eabq6120, 2023.

\bibitem{chate2008modeling}
Hugues Chat{\'e}, Francesco Ginelli, Guillaume Gr{\'e}goire, Fernando Peruani,
  and Franck Raynaud.
\newblock Modeling collective motion: variations on the vicsek model.
\newblock {\em The European Physical Journal B}, 64:451--456, 2008.

\bibitem{helbing2001traffic}
Dirk Helbing.
\newblock Traffic and related self-driven many-particle systems.
\newblock {\em Reviews of modern physics}, 73(4):1067, 2001.

\bibitem{motsch2014heterophilious}
Sebastien Motsch and Eitan Tadmor.
\newblock Heterophilious dynamics enhances consensus.
\newblock {\em SIAM review}, 56(4):577--621, 2014.

\bibitem{mckean1966class}
Henry~P McKean~Jr.
\newblock A class of markov processes associated with nonlinear parabolic
  equations.
\newblock {\em Proceedings of the National Academy of Sciences},
  56(6):1907--1911, 1966.

\bibitem{sznitman1991topics}
Alain-Sol Sznitman.
\newblock Topics in propagation of chaos.
\newblock {\em Ecole d’{\'e}t{\'e} de probabilit{\'e}s de Saint-Flour
  XIX—1989}, 1464:165--251, 1991.

\bibitem{huray2009maxwell}
Paul~G Huray.
\newblock {\em Maxwell's equations}.
\newblock John Wiley \& Sons, 2009.

\bibitem{chen2021solving}
Xiaoli Chen, Liu Yang, Jinqiao Duan, and George~Em Karniadakis.
\newblock Solving inverse stochastic problems from discrete particle
  observations using the fokker--planck equation and physics-informed neural
  networks.
\newblock {\em SIAM Journal on Scientific Computing}, 43(3):B811--B830, 2021.

\bibitem{lu2024weak}
Liwei Lu, Zhijun Zeng, Yan Jiang, Yi~Zhu, and Pipi Hu.
\newblock Weak collocation regression method: Fast reveal hidden stochastic
  dynamics from high-dimensional aggregate data.
\newblock {\em Journal of Computational Physics}, 502:112799, 2024.

\bibitem{kumar2024automating}
Akarsh Kumar, Chris Lu, Louis Kirsch, Yujin Tang, Kenneth~O Stanley, Phillip
  Isola, and David Ha.
\newblock Automating the search for artificial life with foundation models.
\newblock {\em arXiv preprint arXiv:2412.17799}, 2024.

\end{thebibliography}

\end{document}